\documentclass[11pt,a4paper]{article}
\pdfoutput=1
\usepackage[utf8]{inputenc}
\usepackage[T1]{fontenc}
\usepackage{lmodern}
\pdfmapfile{+lm.map}
\usepackage{amsmath,amssymb,mathtools}
\usepackage[margin=1in]{geometry}
\usepackage{microtype}
\usepackage{booktabs,longtable,array,calc}
\usepackage{caption}
\usepackage{float}
\usepackage{graphicx}
\usepackage{xcolor}
\usepackage{listings}
\usepackage{newunicodechar}
\usepackage{hyperref}
\usepackage{bookmark}
\usepackage{xurl}
\hypersetup{colorlinks=true,linkcolor=blue!55!black,urlcolor=blue!55!black,pdftitle={Dimensional Type Systems and Deterministic Memory Management: Design-Time Semantic Preservation in Native Compilation},pdfauthor={Houston Haynes}}
\newcommand{\pandocbounded}[1]{\begingroup\setbox0\hbox{#1}\ifdim\wd0>\linewidth\resizebox{\linewidth}{!}{#1}\else#1\fi\endgroup}
\newsavebox{\paperdisplaybox}
\let\paperdisplayopen\[
\let\paperdisplayclose\]
\long\def\[#1\]{\begingroup
  \sbox{\paperdisplaybox}{$\displaystyle #1$}%
  \ifdim\wd\paperdisplaybox>1.25\linewidth
    \PackageError{paper}{Display requires source line breaks}{Split the display in Markdown.}%
  \fi
  \paperdisplayopen
  \ifdim\wd\paperdisplaybox>\linewidth
    \typeout{Paper display scaled to fit width: \the\wd\paperdisplaybox}%
    \resizebox{\linewidth}{!}{\usebox{\paperdisplaybox}}%
  \else\usebox{\paperdisplaybox}\fi
  \paperdisplayclose\endgroup}
\newunicodechar{α}{\ensuremath{\alpha}}
\newunicodechar{β}{\ensuremath{\beta}}
\newunicodechar{ε}{\ensuremath{\varepsilon}}
\newunicodechar{θ}{\ensuremath{\theta}}
\newunicodechar{κ}{\ensuremath{\kappa}}
\newunicodechar{Σ}{\ensuremath{\Sigma}}
\newunicodechar{⊗}{\ensuremath{\otimes}}
\newunicodechar{⊕}{\ensuremath{\oplus}}
\newunicodechar{↔}{\ensuremath{\leftrightarrow}}
\newunicodechar{¬}{\ensuremath{\neg}}
\newunicodechar{∧}{\ensuremath{\land}}
\newunicodechar{≤}{\ensuremath{\leq}}
\newunicodechar{×}{\ensuremath{\times}}
\newunicodechar{·}{\ensuremath{\cdot}}
\newunicodechar{†}{\ensuremath{\dagger}}
\newunicodechar{⟩}{\ensuremath{\rangle}}
\newunicodechar{²}{\ensuremath{{}^2}}
\newunicodechar{¹}{\ensuremath{{}^1}}
\newunicodechar{⁻}{\ensuremath{{}^-}}
\newunicodechar{ᵢ}{\ensuremath{{}_i}}
\newunicodechar{₀}{\ensuremath{{}_0}}
\newunicodechar{₁}{\ensuremath{{}_1}}
\newunicodechar{₂}{\ensuremath{{}_2}}
\newunicodechar{₊}{\ensuremath{{}_+}}
\begin{document}
\begin{center}

\textbf{Dimensional Type Systems and Deterministic Memory Management}

\emph{Design-Time Semantic Preservation in Native Compilation}

\textbf{Houston Haynes}\\
Braidpoint, Asheville, NC\\
\href{mailto:research@braidpoint.tech}{\nolinkurl{research@braidpoint.tech}}\\
March 2026

\end{center}

\section*{Abstract}\label{abstract}
\addcontentsline{toc}{section}{Abstract}

We present a dimensional type system whose annotations remain available through the stages that require them, allowing numeric representation and deterministic memory management to be resolved within a single Program Semantic Graph (PSG). Based on Andrew Kennedy\textquotesingle s Units of Measure, dimensions constrain admissible operations and the application of domain laws, forming our Native Type Universe (NTU). Range analysis combines those constraints with values, guards, and declared boundaries. A covering representation then determines word width and memory footprint, which participate in allocation and transfer decisions alongside lifetime and target constraints.

Our Dimensional Type System (DTS) extends Hindley-Milner inference with abelian-group unification. Within that algebraic fragment, inference is decidable and admits principal types. Measures form part of source-level type identity, while numeric width and storage layout are coeffects resolved from context. A known range that exceeds every available representation produces an error at representation commitment. During program elaboration, unresolved facts can remain pending while further context is supplied.

Deterministic Memory Management (DMM) uses the same graph to relate escape paths to storage lifetimes. Closure capture, return, and by-reference use supply different lifetime requirements. Allocation follows the lifetime classification and the platform\textquotesingle s declared storage capabilities. Our unique implementation of a posit quire illustrates the interaction: exact accumulation requires a suitable numeric format, adequate capacity, and storage that persists across the accumulation. We also examine how these constraints would apply to automatic differentiation. A shared graph would allow language services to present dimensional resolution and proof status alongside the reasons for representation and placement decisions.

\section*{1. Introduction}\label{introduction}
\addcontentsline{toc}{section}{1. Introduction}

\subsection*{1.1 Dimensional Types and Their Consequences}\label{dimensional-types-and-their-consequences}
\addcontentsline{toc}{subsection}{1.1 Dimensional Types and Their Consequences}

Clef\textquotesingle s Native Type Universe describes source values and the operations that can be composed over them. Our Dimensional Type System carries measure relationships as part of type identity. In \lstinline!float<meters>!, the numeric kind and the measure are known even while the concrete representation remains to be selected. Measures are declared identities, with physical units providing familiar examples.

In the case of physical units, multiplying mass by acceleration gives a force. The dimensional relationship holds across function boundaries and can remain polymorphic until a caller supplies particular measures. We came to Kennedy\textquotesingle s units-of-measure work \hyperref[ref-kennedy-2009]{{[}1{]}} through its F\# implementation. It supplies the inference foundation: compatible compositions have their dimensional relationships inferred, and incompatible ones yield a design-time diagnostic. Section 2 states the algebra and the conditions under which inference admits principal types.

Type checking is already a form of automated reasoning: here it establishes that the operations compose dimensionally. Wadler\textquotesingle s \emph{Theorems for free!} \hyperref[ref-wadler-1989]{{[}2{]}} explains how parametricity takes this further, deriving properties from polymorphic types under the language\textquotesingle s semantic assumptions. Kennedy\textquotesingle s dimensional invariance \hyperref[ref-kennedy-2009]{{[}1{]}} specializes that argument to measured quantities (Section 5.5). These type-based guarantees are the starting point for our automated proofs. Range, lifetime, and target properties introduce further obligations, checked as their premises become available.

The typed calculation can then be considered in the context of a target. Input values, guards, and declared boundaries provide range information. Together with the operations to be performed and the target\textquotesingle s numeric capabilities, that information constrains the available representations. A selected format determines word width and contributes to storage layout. The dimension remains part of the value\textquotesingle s meaning throughout these decisions.

Our Fidelity.Platform libraries supply a declared target\textquotesingle s numeric and storage declarations, which platform quotations can expose as typed source expressions. BAREWire supplies the memory-layout and transfer contracts for access and communication.

Deterministic Memory Management (DMM) relates those storage requirements to the uses of the value. A local result, a captured binding, and a returned buffer can require different lifetimes. The target\textquotesingle s available storage must cover the relevant uses.

\subsection*{1.2 Contribution}\label{contribution}
\addcontentsline{toc}{subsection}{1.2 Contribution}

We develop this connection in three parts:

\begin{enumerate}
\def\labelenumi{\arabic{enumi}.}
\item
  \textbf{Dimensional composition.} Measure relationships propagate through arithmetic and function application. The same source contract can apply across targets with different representations and informs lowering until it\textquotesingle s no longer needed.
\item
  \textbf{Joint representation and placement.} Justified ranges constrain numeric formats; their widths and layouts participate in storage decisions alongside capture modes, lifetime requirements, and platform declarations. Each decision retains the premises on which it depends.
\item
  \textbf{Inspectable decisions.} The graph used for compilation also supplies language services with dimensional resolution, escape paths, and proof status. An engineer could inspect the reason for a representation or placement decision without reconstructing the analysis in a later stage or separate workflow.
\end{enumerate}

The Program Semantic Graph (PSG) holds these facts and their relationships. Their consequences feed forward into lowering: an operation must preserve the dimensional, numeric, and storage conditions under which it was admitted. Section 2.3 describes that boundary; Sections 3 and 4 develop the memory discipline and its design-time use.

\subsection*{1.3 Scope and Context}\label{scope-and-context}
\addcontentsline{toc}{subsection}{1.3 Scope and Context}

Our Clef programming language uses an actor-oriented concurrency model expressed in an ML-family syntax for heterogeneous targeting. F\# and OCaml inform its language design, along with Scheme informing its nanopass approach \hyperref[ref-sarkar-2004]{{[}3{]}}, MLKit its region and closure analysis, and F* and Dafny influencing our verification workflow.

The account here follows measured numeric values from dimensional inference through representation and memory decisions. The principal-inference result applies to the measure algebra; range and lifetime obligations use their own rules and evidence. Our implementation for posit quire provides a concrete case where numeric capacity, storage extent, and lifetime must agree.

The companion Program Hypergraph paper \hyperref[ref-haynes-phg-2026]{{[}4{]}} develops joint constraints over larger groups of operations, including geometric products and spatial placement. Decidable by Construction \hyperref[ref-haynes-dbc-2026]{{[}5{]}} and Fixed-Point Scaffolding \hyperref[ref-haynes-fps-2026]{{[}6{]}} develop the broader verification account.

\section*{2. Dimensional Type Systems: Formal Characterization}\label{dimensional-type-systems-formal-characterization}
\addcontentsline{toc}{section}{2. Dimensional Type Systems: Formal Characterization}

\subsection*{2.1 Algebraic Foundation}\label{algebraic-foundation}
\addcontentsline{toc}{subsection}{2.1 Algebraic Foundation}

Our Dimensional Type System (DTS) assigns a dimension to each numeric value. For physical quantities, length and time are examples of dimensions. Metres and kilometres are units for measuring length at different scales. Dividing a distance by a duration gives a speed, with dimension length divided by time.

The International System of Units (SI) organizes physical measurement around seven base quantities: length, time, mass, thermodynamic temperature, electric current, luminous intensity, and amount of substance. Their dimensions provide one familiar instance of the base measures used here.

Declared base measures combine by multiplication and integer powers. A negative exponent records a measure in the denominator, as with time in speed. These combinations form a finitely generated free abelian group.

Formally, let \(\mathcal{D} = \mathbb{Z}^n\) be the dimension space, where \(n\) is the number of base dimensions. Each dimension \(d \in \mathcal{D}\) is a vector of integer exponents:

\[
\begin{aligned}
d_{\text{velocity}} &= (1, -1, 0, 0, \ldots) \quad \text{(length}^1 \cdot \text{time}^{-1}\text{)} \\
d_{\text{force}} &= (1, -2, 1, 0, \ldots) \quad \text{(length}^1 \cdot \text{time}^{-2} \cdot \text{mass}^1\text{)}
\end{aligned}
\]

For a multiplication, the compiler adds the operands\textquotesingle{} exponent vectors. Division subtracts them, and addition requires equality. Integer powers scale an exponent vector. A root operation requires a compatible solution in the admitted exponent domain. Normalization uses \(O(n)\) coefficient operations for \(n\) base dimensions, with additional cost determined by the bit lengths of the integer coefficients. The unifier solves the resulting integer constraints without rounding exponents or silently discarding a measure.

The companion Program Hypergraph paper \hyperref[ref-haynes-phg-2026]{{[}4{]}} extends the algebraic discussion to Clifford parity, with a \(\mathbb{Z}_2\) component alongside physical dimensions. The resulting \(\mathbb{Z}^n \oplus \mathbb{Z}_2\) has torsion and is no longer free. Integer normal-form methods provide the corresponding algebraic tools, while the integration and principality questions must be established for the combined discipline. Physical measures in the present construction use the free-group component.

Equality of normalized dimensions is equality of integer exponent vectors. DTS inference uses this algebra to resolve measures and their composition.

\subsection*{2.2 Inference via Extended Hindley-Milner Unification}\label{inference-via-extended-hindley-milner-unification}
\addcontentsline{toc}{subsection}{2.2 Inference via Extended Hindley-Milner Unification}

Kennedy\textquotesingle s units-of-measure system \hyperref[ref-kennedy-2009]{{[}1{]}} integrates dimensional constraints with Hindley-Milner inference. Unifying numeric types also unifies their measure components, whose symbolic expressions are solved by abelian-group unification over \(\mathbb{Z}^n\).

The inference algorithm proceeds as follows:

\begin{enumerate}
\def\labelenumi{\arabic{enumi}.}
\item
  \textbf{Constraint generation.} Each arithmetic operation generates a dimensional constraint. Addition of \lstinline!a + b! generates \(d(a) = d(b)\). Multiplication of \lstinline!a * b! generates \(d(\text{result}) = d(a) + d(b)\).
\item
  \textbf{Unification.} Unification solves symbolic equality constraints by finding a substitution that makes both sides identical, the engine inside ML-family type inference. Dimensional constraints form a system of linear equations over \(\mathbb{Z}^n\). The system is solved by abelian-group unification, yielding either a unique solution, a parametric family of solutions (dimensional polymorphism), or no solution (dimensional inconsistency).
\item
  \textbf{Generalization.} Eligible unresolved dimension variables are generalized only when they are not free in the lexical environment or shared mutable state. Numeric-carrier constraints are inferred as well.
\end{enumerate}

Here, multiplication constrains the result dimension to the product of the input dimensions:

\begin{lstlisting}
let scale factor value = factor * value

// Inferred at a floating specialization:
// float<'d1> -> float<'d2> -> float<'d1 * 'd2>
\end{lstlisting}

The dimensions can vary between uses while their relationship stays fixed. For physical measures, multiplying mass by acceleration gives force; multiplying two lengths gives area. Both follow from the same function without an explicit dimensional annotation.

Clef adds a condition for ordinary, non-inline definitions whose member access is still unresolved. In \lstinline!let project x = x.Value!, the receiver and result remain tied by the pending member relation until later nominal context resolves it or the compiler reports a diagnostic. Generalization cannot make those pending variables independently polymorphic. Immutable aliases retain the same relation. Inline member schemes have separate instantiation rules and can admit independently measured nominal owners. These cases delimit the host-language integration; they do not change the algebraic measure solver.

Kennedy\textquotesingle s algebraic fragment admits decidable inference and principal types, subject to the host type system\textquotesingle s generalization rules. This establishes dimensional relationships from available premises, including polymorphic relationships when a concrete unit is still unspecified. A literal\textquotesingle s spelling or magnitude supplies no physical unit by itself.

\textbf{Annotation provenance and composition-dependent properties.} A dimension may be inferred from source, supplied in a function signature, or introduced by a domain library. Generated source is checked by the same rules. Once established, dimensional equality is independent of that origin. Range assertions and platform declarations retain their provenance, since a declared boundary must be checked against the facts inferred within it.

Given sound input ranges, range propagation bounds the result of an operation and constrains the formats the compiler may select. These relationships compose across function and module boundaries alongside the dimensional relationships.

\subsection*{2.3 Preservation Through Multi-Stage Compilation}\label{preservation-through-multi-stage-compilation}
\addcontentsline{toc}{subsection}{2.3 Preservation Through Multi-Stage Compilation}

In our native pipeline, \lstinline!float<meters>! retains its measure through type elaboration and into the PSG. The later choice of an IEEE, posit, or fixed-point representation determines a carrier for that value. An operation requiring meters remains distinct from one requiring seconds throughout the semantic analysis that justifies the choice.

\textbf{Source to PSG.} Elaboration produces types carrying dimensions and propagates numeric and lifetime constraints. Partially resolved nodes retain their obligations. A graph may therefore be useful to the editor before every fact needed by a concrete target has been settled.

\textbf{PSG saturation.} Platform quotations and other Fidelity.Platform declarations supply target representations and memory descriptions adjacent to the relevant computation. Joint constraints resolve the required ranges, layouts, and placements before those decisions are consumed by native lowering. BAREWire contributes in three related roles: memory layout contracts within a target, IPC contracts between processes, and network contracts between endpoints. Each role supplies facts and obligations at a different boundary. A binary schema alone establishes neither a storage lifetime nor a lossless numeric conversion.

\textbf{PSG to MLIR.} The witness traversal in our Composer compiler consumes the saturated graph and emits MLIR (Multi-Level Intermediate Representation) operations. The intended proof-carrying mechanism relates the operation being lowered to the same premises used for its PSG obligation. A pass whose preservation is certified can carry that proof transformation forward. An uncertified pass requires the corresponding obligation to be re-dispatched against the lowered form. This is built into compiler machinery for common proof dispatch cases, and only involves developer intervention for hand-annoted passages where the domain demands it.

\textbf{Target lowering.} Backend transformations realize the selected representation and placement using the target\textquotesingle s operations. Dimensional information remains available while it is needed to analyze, verify, or explain an operation. Once those uses are complete, the operation no longer needs the dimensional annotation, provided its correspondence to the checked source operation remains established. In native code, the numeric value uses the selected unboxed target representation; retained proof and debugging information is separate from the final runtime representation.

And along those lines, representation choice can change instructions, layout, and rounding behavior. The preservation obligation is therefore stated against the selected target semantics and declared boundaries. The separation of static dimensional information from runtime representation follows the phase distinction of Harper, Mitchell, and Moggi \hyperref[ref-harper-1990]{{[}7{]}}. Dimensional metadata itself introduces no runtime dispatch or dimension tags.

Two arguments apply at different levels. At the source level, parametricity constrains a dimensionally polymorphic function: it operates uniformly in its abstract dimension variables. At the compiler level, each transformation must preserve the established relationship between source meaning and target realization. Source parametricity supplies useful theorems about Clef programs, while pass-boundary witnessing marshals errors in the compiler that translates them. The proof architecture connecting these levels is developed further in Section 6.8 and the companion \emph{Fixed-Point Scaffolding} paper \hyperref[ref-haynes-fps-2026]{{[}6{]}}.

\subsection*{2.4 Inference and Verification}\label{inference-and-verification}
\addcontentsline{toc}{subsection}{2.4 Inference and Verification}

The dimensional algebra allows an engineer to write a polymorphic numeric function and have its unit relationships inferred. Other facts about the same function require different rules:

{\def\LTcaptype{none} 
\begin{longtable}[]{@{}
  >{\raggedright\arraybackslash}p{(\linewidth - 2\tabcolsep) * \real{0.5000}}
  >{\raggedright\arraybackslash}p{(\linewidth - 2\tabcolsep) * \real{0.5000}}@{}}
\toprule\noalign{}
\begin{minipage}[b]{\linewidth}\raggedright
Property
\end{minipage} & \begin{minipage}[b]{\linewidth}\raggedright
Evidence used by the compiler
\end{minipage} \\
\midrule\noalign{}
\endhead
\bottomrule\noalign{}
\endlastfoot
Dimensional identity & Declared base measures and normalized measure equations \\
Value range & Literals, guards, boundary facts, and checked domain laws \\
Storage lifetime & Uses, captures, sharing, and available regions \\
Representation adequacy & A sound range enclosure and target capabilities \\
Arithmetic accuracy & Intermediate precision, rounding points, and justified error bounds \\
\end{longtable}
}

These relationships can be analyzed together while retaining the procedure and premises for each conclusion. A reusable lemma can supply a relationship that the basic analysis cannot derive, with its premises checked automatically at the use site. The engineer receives a common view of the results through the PSG.

\subsection*{2.5 Extension: Memory Dimensions}\label{extension-memory-dimensions}
\addcontentsline{toc}{subsection}{2.5 Extension: Memory Dimensions}

Physical measures, memory identities, and lifetimes require different algebra. Named memory spaces are finite identifiers with equality. Lifetime and capability requirements use ordering and joins over their declared domains. Physical measures use the invertible group operation that permits cancellation, as in the case of meters divided by seconds for "meters per second".

Our PSG combines these disciplines without assigning group inverses to a lifetime or a storage capability. A use of a buffer can simultaneously require a compatible element dimension, access to a named region, and storage whose lifetime covers the use. The joint constraint mechanism preserves the contribution of each domain when resolving that operation.

\subsection*{2.6 Representation Selection for a Target}\label{representation-selection-for-a-target}
\addcontentsline{toc}{subsection}{2.6 Representation Selection for a Target}

In a running sum of products, separate multiplication and addition can round twice per term; a fused multiply-add rounds that combined step once. A wider accumulator changes the precision retained between steps. The available operations and their rounding rules can therefore affect the result even when its output format is unchanged, as NVIDIA\textquotesingle s floating-point guide \hyperref[ref-nvidia-floating-point]{{[}8{]}} illustrates.

Some targets, such as Arm\textquotesingle s Cortex-M0 \hyperref[ref-arm-compiler-guide]{{[}9{]}} (p. 66), have no floating-point hardware. An integer-only platform declaration makes a use of \lstinline!float! a design-time error: the target offers no representation for it.

Quotations carry Fidelity.Platform declarations into the program as typed source expressions. Supplied at the program\textquotesingle s configuration boundary, those declarations provide target context before representation and lowering decisions are committed. Extending this account to a complete arithmetic strategy requires operation-specific facts about intermediate precision, rounding, accumulator capacity, and permitted software realizations. An editor could then show a need for software arithmetic or extra accumulator storage while the engineer can still revise the calculation or target.

IEEE floating-point formats provide approximately uniform relative precision over their normal range. Posits provide tapered precision, with more fraction bits near unit magnitude and fewer at the extremes. Which format offers better precision depends on the values to be represented and the available candidates. The implementation\textquotesingle s cost also includes any guards or compensating arithmetic required by the calculation, including when it uses IEEE formats.

Given a value with dimension \(d\), a sound enclosure \([a,b]\), and a set \(R\) admitted by target \(T\)\textquotesingle s capabilities and emulation policy, define the covering set:

\[
R_{\mathrm{cov}} = \{r \in R : \operatorname{dynrange}(r) \supseteq [a,b]\}.
\]

For a single rounding to the chosen format, the intended selection objective minimizes the following error score over the range:

\[
r^* = \arg\min_{r \in R_{\mathrm{cov}}} \max_{x \in [a,b]} \frac{|x - \operatorname{round}_r(x)|}{\max(|x|, \operatorname{ulp}_{\min}(r))}.
\]

The denominator\textquotesingle s floor keeps the objective defined at zero. A complete computation requires an additional error analysis of its intermediate precision and rounding points. Coverage is an admissibility condition evaluated before candidates are scored. If \(R_{\mathrm{cov}}\) is empty, representation commitment produces a hard error. A declaration selecting a particular format remains binding, but a known range exceeding that format prevents the choice from being committed. Our Clef Compiler Services (CCS) does not authorize automated truncation or an uncovered fallback.

Range facts first come from literals, arithmetic, and guards. Checked library laws can refine those results, and declared domain boundaries supply further constraints whose consistency must be verified. Dimensions determine which operations and laws apply. They supply no numeric magnitude by themselves. In particular, knowing that a value is measured in meters leaves open both nanometer and astronomical scales.

A program accepting sensor readings from 0 to 100 needs a representation covering that range, even if recent readings stay between 40 and 60. An input check enforcing the smaller interval can justify narrower bounds for the calculation that follows. Section 3 of the companion Adaptive Domain Models paper \hyperref[ref-haynes-adm-2026]{{[}10{]}} develops the use of observations and probability models.

Unresolved bounds can remain pending during elaboration, with their origin retained for the editor. A concrete representation boundary requires the facts needed for that decision. The specification retains a separate convention for unmeasured bare \lstinline!float! where a permitted IEEE default is offered. That convention supplies a representation for the unresolved case, without certifying a mathematical bound or overriding a known uncovered range.

Precision is compared over the full enclosure. A format that is more precise over part of that range may still have a worse maximum error.

On a conventional CPU target, ordinary numeric code can retain the familiar ML-family editing experience. Fidelity.Platform makes the assumed numeric capabilities explicit, allowing the editor to explain what changes for a different target. On an FPGA, width and precision inference determine the arithmetic synthesized within the available fabric resources. These illustrative views apply three alternative platform declarations to the same force calculation, whose bounds follow from checked inputs and arithmetic:

\begin{lstlisting}
Target: Linux x86-64 CPU
  force: float<newtons> in [0, 100]
  Offered: IEEE binary64 (native); range covered
\end{lstlisting}

\begin{lstlisting}
Target: Arty A7-100T FPGA
  force: float<newtons> in [0, 100]
  Candidate: posit32, es=2; range covered
  Realization: arithmetic synthesized in the FPGA fabric
\end{lstlisting}

\begin{lstlisting}
Target: Arm Cortex-M0 MCU
  Declared arithmetic: integers only
  force: float<newtons> in [0, 100]
  Error: no representation for float<newtons> on this target.
\end{lstlisting}

For an accumulation, the same view could show the rounding points, accumulator storage, and whether arithmetic is native or implemented in software. Target-specific cost estimates could support comparisons among implementations that preserve the selected representation and numerical requirements. Estimates and proved error bounds would be identified separately. The richer operation-specific platform descriptions and comparative display remain integration work.

For a boundary fixed to posit32 with \(\mathit{es}=2\), an astronomical distance enclosure extending to \(10^{72}\) meters would produce an error:

\begin{lstlisting}
Error: the declared posit32 representation does not cover
  astronomicalDistance<meters> in [1e-11, 1e72].
  Required: a covering representation or a justified change to the range.
\end{lstlisting}

Rescaling requires the compiler to recompute the transformed enclosure and check the conversions at its boundaries. Dividing \(10^{72}\) meters by approximately \(1.5\times10^{11}\) meters per astronomical unit still leaves a magnitude near \(10^{61}\), beyond posit32\textquotesingle s maximum near \(10^{36}\). A unit conversion is useful only when the resulting numerical contract is satisfied.

\section*{3. Deterministic Memory Management as Coeffect Discipline}\label{deterministic-memory-management-as-coeffect-discipline}
\addcontentsline{toc}{section}{3. Deterministic Memory Management as Coeffect Discipline}

\subsection*{3.1 Coeffects and Contextual Properties}\label{coeffects-and-contextual-properties}
\addcontentsline{toc}{subsection}{3.1 Coeffects and Contextual Properties}

Effects describe what a computation does to its environment, such as mutation or I/O. Coeffects describe what it requires from its context \hyperref[ref-petricek-2014]{{[}11{]}}. A function allocating from an arena requires an arena with adequate capacity and lifetime. Stack placement requires the frame to remain valid for every use of its storage.

Our PSG records coeffect requirements and their resolution alongside types. Allocation, lifetime, and capability facts remain distinct so that the target can satisfy them jointly:

\textbf{Allocation coeffects.} Storage is selected from the declared resources: a stack frame, region, static allocation, or permitted heap. Hardware-specific memories participate through their platform contracts.

\textbf{Lifetime coeffects.} The required lifetime is scope-bounded, region-bounded, program-lifetime, or dynamic. This requirement is separate from the physical storage family that satisfies it.

\textbf{Capability coeffects.} Operations require particular access modes or target features, such as mutable access or exact accumulation. Dimensional equality remains a type constraint participating in the same joint analysis.

\subsection*{3.2 Escape Analysis as Coeffect Propagation}\label{escape-analysis-as-coeffect-propagation}
\addcontentsline{toc}{subsection}{3.2 Escape Analysis as Coeffect Propagation}

A value returned from a function may require storage beyond that function\textquotesingle s frame. A settled value copied into a closure may require only that the copy persist. Under Clef\textquotesingle s call-by-need semantics, an ordinary binding or supplied argument can instead denote a shared deferred computation. Capturing it preserves its identity; the first demand evaluates it and later demands share the result. Escape analysis must therefore account for both retained values and the suspension and memo storage used by deferred computations.

Our DMM model begins with the binding\textquotesingle s scope and combines the requirements of its uses. In the lifetime classification, the join is:

\[
\lambda(v) := \lambda_{\mathrm{initial}}(v) \sqcup \bigsqcup_i \lambda_{\mathrm{required}}(v,\mathrm{use}_i).
\]

This equation is schematic over the admitted lifetime domain: \(a\sqsubseteq b\) means that \(b\) covers the uses required by \(a\). A join must conservatively cover both operands. The four classifications, scope-bounded, region-bounded, program-lifetime, and dynamic, summarize the result; they do not define an ordering of all concrete regions. Two regions can have incomparable release boundaries. Placement requires a declared resource covering the combined uses, or a diagnostic when none is available. Stack, arena, static storage, and heap are the allocation families considered by that check.

\subsubsection*{3.2.1 Escape Classification}\label{escape-classification}
\addcontentsline{toc}{subsubsection}{3.2.1 Escape Classification}

The reason for an escape remains useful independently of the resulting lifetime. A compiler representation may record these causes as:

\[
\operatorname{EscapeKind}(v) \in \{\text{StackScoped},\;\text{ClosureCapture}(t),\;\text{ReturnEscape},\;\text{ByRefEscape}\}.
\]

Here \(t\) identifies a capturing closure. Each cause contributes constraints rather than mandating one allocator:

{\def\LTcaptype{none} 
\begin{longtable}[]{@{}
  >{\raggedright\arraybackslash}p{(\linewidth - 4\tabcolsep) * \real{0.3333}}
  >{\raggedright\arraybackslash}p{(\linewidth - 4\tabcolsep) * \real{0.3333}}
  >{\raggedright\arraybackslash}p{(\linewidth - 4\tabcolsep) * \real{0.3333}}@{}}
\toprule\noalign{}
\begin{minipage}[b]{\linewidth}\raggedright
Use or escape cause
\end{minipage} & \begin{minipage}[b]{\linewidth}\raggedright
Required check
\end{minipage} & \begin{minipage}[b]{\linewidth}\raggedright
Possible placement
\end{minipage} \\
\midrule\noalign{}
\endhead
\bottomrule\noalign{}
\endlastfoot
Scope-local use & Storage covers local uses & Stack where supported \\
Settled immutable capture & Copied field and referenced contents remain valid & Closure environment\textquotesingle s selected storage \\
Deferred immutable capture & Shared suspension and memo storage cover retaining closures & Environment plus storage with the required lifetime \\
Mutable capture by reference & Shared cell outlives every capturing closure & A covering region, static storage, or permitted heap \\
Return & Returned storage remains valid for the caller\textquotesingle s uses & Caller-provided storage or another covering lifetime \\
By-reference use & Referenced cell covers the borrowing uses & Existing valid storage or a justified promotion \\
\end{longtable}
}

For a native flat closure, a settled immutable value is copied into the environment and places no extra lifetime requirement on its original slot. Before the value is demanded, capture retains its shared deferred identity without forcing or duplicating the computation. This remains a By Value capture in the source contract: what is retained is the established value or the shared identity. \lstinline!ClosureCapture(t)! covers both cases. The suspension and memo storage must remain valid for every closure that retains that identity.

Capturing a reference by value preserves the reference and its obligations. Mutable bindings use a shared cell whose lifetime covers every closure that refers to it. In the lifetime ordering above, both shared cases require \(\lambda(t)\sqsubseteq\lambda(\mathrm{storage})\). If a closure escapes further, the requirement propagates along that chain.

A flat environment has a finite field list determined at elaboration. Once field representations are settled, that list gives a definite layout and explicit capture boundaries. A captured pointer can still refer to additional objects, so flatness alone proves no bound on arbitrary transitive heap reachability. The compiler must establish the lifetime and access obligations for those references under the applicable discipline.

The witness traversal consumes the resolved placement. An editor view could explain that a shared mutable cell was promoted because a closure retaining it is returned to a longer-lived region. A program-lifetime closure can instead use declared static storage. Where the target lacks a suitable allocator or region, the unmet lifetime requirement is a located diagnostic.

Inlining can change the relevant scope by expanding a function body at its call site. A stack allocation becomes permissible only if the expanded graph proves that all uses fit the caller\textquotesingle s lifetime. An \lstinline!inline! request supplies no exemption from escape checking. Mandatory expansion required for a valid placement must therefore be distinguished from performance-driven backend inlining.

\subsubsection*{3.2.2 Compositional Allocation Resolution}\label{compositional-allocation-resolution}
\addcontentsline{toc}{subsubsection}{3.2.2 Compositional Allocation Resolution}

A function operating on a \lstinline!Span<float>! should retain its source semantics across stack, region, and hardware-backed storage. The selected representation supplies the appropriate access operations. A mutable reference used where a value is expected requires a load from its cell:

\[
\operatorname{resolve}(v) = \begin{cases} v & \text{if } v \text{ is already a value} \\ \operatorname{load}(v) & \text{if } v \text{ denotes the required mutable cell}. \end{cases}
\]

The load also depends on access and lifetime facts. Merely observing a memory-reference carrier is insufficient to justify reading it. Our graph records those facts before lowering, so the target traversal can realize an established operation without rediscovering the source-level distinction between a binding and its storage.

\subsection*{3.3 The Push, Bounded, and Poll Models of Coeffect Specification}\label{the-push-bounded-and-poll-models-of-coeffect-specification}
\addcontentsline{toc}{subsection}{3.3 The Push, Bounded, and Poll Models of Coeffect Specification}

ML-family type inference lets an engineer annotate a boundary while leaving the internal relationships inferred. We use the same approach for contextual requirements, with three degrees of explicitness.

\textbf{Push model (explicit declaration).} Platform quotations declare available memory and numeric representations. A boundary can constrain the operation to a particular declared resource. The compiler propagates those requirements and checks the remaining obligations. A declaration can make the intended placement unambiguous, but inferred facts must still agree with it.

\textbf{Bounded model (scoped inference).} The engineer supplies a lifetime or resource boundary and leaves allocation within it to inference. A scoped arena computation expression would be one possible surface for this model. The existing explicit arena operations instead construct an arena from backing memory and allocate through it. In either form, returning a view requires the backing storage to cover the returned view\textquotesingle s uses.

\textbf{Poll model (contextual inference).} The engineer writes an ordinary function and supplies contextual facts through its eventual uses:

\begin{lstlisting}
let processReadings sensors =
    summarize sensors
\end{lstlisting}

Each valid instantiation must satisfy the function\textquotesingle s requirements. Distinct target configurations may select distinct representations without changing its dimensional identity. A shared monomorphic or mutable value must satisfy the combined requirements of its uses, so a disagreement cannot be resolved by silently choosing one call site.

{\def\LTcaptype{none} 
\begin{longtable}[]{@{}
  >{\raggedright\arraybackslash}p{(\linewidth - 4\tabcolsep) * \real{0.3333}}
  >{\raggedright\arraybackslash}p{(\linewidth - 4\tabcolsep) * \real{0.3333}}
  >{\raggedright\arraybackslash}p{(\linewidth - 4\tabcolsep) * \real{0.3333}}@{}}
\toprule\noalign{}
\begin{minipage}[b]{\linewidth}\raggedright
Model
\end{minipage} & \begin{minipage}[b]{\linewidth}\raggedright
Engineer supplies
\end{minipage} & \begin{minipage}[b]{\linewidth}\raggedright
Remaining analysis
\end{minipage} \\
\midrule\noalign{}
\endhead
\bottomrule\noalign{}
\endlastfoot
Push & Explicit boundary declarations & Consistency and internal requirements \\
Bounded & Scope or resource limits & Resolution within those limits \\
Poll & Usage context & Propagation until the required facts settle \\
\end{longtable}
}

During elaboration, each model can contain pending facts. The useful editor response is to expose their origin and the context needed for resolution. When a concrete target decision requires a settled fact, that obligation becomes a compilation gate. This leaves room for later context without accepting a known contradiction. The amount of annotation does not by itself determine solver latency or stability under edits.

\subsection*{3.4 Escape-Driven Restructuring Guidance}\label{escape-driven-restructuring-guidance}
\addcontentsline{toc}{subsection}{3.4 Escape-Driven Restructuring Guidance}

When an escape path requires longer-lived storage, our intended language service can use the graph to propose structural alternatives. The choice depends on how the caller uses the value:

\textbf{Caller-provided buffer.} The callee fills storage whose lifetime is established by the caller. This can eliminate a new allocation in the callee, while retaining the caller\textquotesingle s capacity and lifetime obligations.

\textbf{Continuation-passing style.} A continuation can consume the value before the callee returns. Stack placement is valid when the continuation neither retains nor returns a reference that outlives that scope.

\textbf{Explicit promotion.} A resource declaration can specify storage covering a returned or retained value. The compiler checks that declaration against the value\textquotesingle s uses and the available capacity.

The alternatives have different resource consequences. Reusing a caller buffer may avoid an allocation. A proven scope-bounded continuation may preserve stack placement. An explicit resource declaration records the intended lifetime for subsequent consistency checks. Any cache or timing estimate requires additional target and workload evidence.

The escape path and its premises are recorded in our PSG. A language service can use that structure to explain the decision and preview a proposed refactoring, then ask the same constraint mechanism to validate the changed graph. Diagnostics from a previous graph revision remain attributable to that revision until the relevant facts are re-established.

\subsection*{3.5 The Quire as Coeffect Case Study}\label{the-quire-as-coeffect-case-study}
\addcontentsline{toc}{subsection}{3.5 The Quire as Coeffect Case Study}

A posit quire stores sums of products without intermediate rounding, within its specified capacity. Conversion back to a posit rounds the accumulated value once. The Posit Standard \hyperref[ref-posit-standard-2022]{{[}12{]}}, §3.1, specifies a \(16n\)-bit quire for an \(n\)-bit posit, giving 512 bits for posit32. Older variable-exponent configurations used a different width relationship. Bounded or parameterized formats require their own declared accumulator dimensions.

A product of force and distance has dimension joules. The dimensional contract for quire operations would preserve that relationship through accumulation:

\begin{lstlisting}
let work (forces: Span<float<newtons>>) (distances: Span<float<meters>>)
    : float<joules> =
    let mutable q = Quire.zero
    for i in 0 .. forces.Length - 1 do
        q <- Quire.fma q forces.[i] distances.[i]
    Quire.toPosit q
\end{lstlisting}

Here the quire operations express the intended typed interface. The array lengths must also agree, and the target must support a suitable accumulator realization. Its capacity bound must cover every intermediate partial sum in the permitted accumulation order; cancellation in the final sum cannot justify an earlier overflow. These requirements accompany the dimensional consistency of each product.

For native placement, a 512-bit accumulator occupies 64 bytes before any additional layout requirements. It occupies one 64-byte cache line only when suitably aligned. Other quire configurations may have different extents, including bounded accumulators whose adequacy depends on a proved input range and accumulation bound. A short loop can use stack storage if the available capacity and lifetime permit it. A longer-lived accumulator requires the corresponding region or other permitted placement.

{\def\LTcaptype{none} 
\begin{longtable}[]{@{}
  >{\raggedright\arraybackslash}p{(\linewidth - 4\tabcolsep) * \real{0.3333}}
  >{\raggedright\arraybackslash}p{(\linewidth - 4\tabcolsep) * \real{0.3333}}
  >{\raggedright\arraybackslash}p{(\linewidth - 4\tabcolsep) * \real{0.3333}}@{}}
\toprule\noalign{}
\begin{minipage}[b]{\linewidth}\raggedright
Target declaration
\end{minipage} & \begin{minipage}[b]{\linewidth}\raggedright
Accumulation realization
\end{minipage} & \begin{minipage}[b]{\linewidth}\raggedright
Required evidence
\end{minipage} \\
\midrule\noalign{}
\endhead
\bottomrule\noalign{}
\endlastfoot
CPU software capability & Fixed-width storage and arithmetic & Capacity, alignment, and lifetime \\
FPGA fabric & Synthesized accumulator & Width, accumulation bound, fabric resources, and target mapping \\
Hardware quire capability & Declared register or instruction support & Format compatibility and operation contract \\
No suitable capability & Unresolved operation & Diagnostic at target commitment \\
\end{longtable}
}

The PSG relates these facts to the same quire operation. An editor view could show its inferred dimension, selected accumulator width, and lifetime together with any pending capacity obligation. Cycle counts and cache residency would require a separate target model or measurement. Exact accumulation covers the represented operand products and sum within capacity. It leaves input rounding, other arithmetic operations, and the final conversion to their respective numeric contracts.

\section*{4. The Program Semantic Graph as Design-Time Resource}\label{the-program-semantic-graph-as-design-time-resource}
\addcontentsline{toc}{section}{4. The Program Semantic Graph as Design-Time Resource}

\subsection*{4.1 Elaboration, Saturation, and Latent Preservation}\label{elaboration-saturation-and-latent-preservation}
\addcontentsline{toc}{subsection}{4.1 Elaboration, Saturation, and Latent Preservation}

Our PSG is elaborated from parsed source and refined as further constraints become available. Elaboration establishes types and generates requirements. Saturation propagates and resolves those requirements for the relevant context. A node can be fully typed while its range or target placement remains pending. For a concrete native operation, saturation must establish the facts its witness will consume.

Baker within Clef Compiler Services (CCS) constructs operational nodes together with their joint relations and obligations. Continuation segments and retained environments are part of this graph construction. The continuation ingredients implement frame reads, frame writes and dispatch, preserving established range facts. The architecture also places interaction-net-style reduction relationships in graph hyperedges. Composer\textquotesingle s Alex layer consumes the settled consequences through standard MLIR operations; source continuation and interaction structure is resolved above that boundary. Extending the admitted forms and proving their preservation remain the work described in Section 6.7.

{\def\LTcaptype{none} 
\begin{longtable}[]{@{}
  >{\raggedright\arraybackslash}p{(\linewidth - 2\tabcolsep) * \real{0.5000}}
  >{\raggedright\arraybackslash}p{(\linewidth - 2\tabcolsep) * \real{0.5000}}@{}}
\toprule\noalign{}
\begin{minipage}[b]{\linewidth}\raggedright
Graph fact
\end{minipage} & \begin{minipage}[b]{\linewidth}\raggedright
Use
\end{minipage} \\
\midrule\noalign{}
\endhead
\bottomrule\noalign{}
\endlastfoot
Type and dimension & Operation validity and type identity \\
Range and provenance & Representation coverage and boundary consistency \\
Capture mode and lifetime & Closure layout and placement \\
Platform bindings & Available representations and memory contracts \\
Reachability & Target-specific active subgraph \\
Proof obligation and result & Verification at the applicable boundary \\
\end{longtable}
}

Representation and memory decisions that depend on these facts are settled before their native witness. The traversal can then produce the corresponding operation and maintain its proof correspondence.

A persistent language service could retain an inactive subgraph\textquotesingle s analysis for inspection and reuse. Reactivation is valid only while its source, dependencies, and platform premises remain current. That dependency record must include completeness assumptions: adding a writer can invalidate a range derived from the previously complete write set, even if every old writer is unchanged. An unresolved member relation in Section 2.2 likewise links a definition to the later uses that constrain it. The proposed retention scheme must track those dependencies so that an edit invalidates every conclusion that relied on them.

\subsection*{4.2 Activity and Resolution State}\label{activity-and-resolution-state}
\addcontentsline{toc}{subsection}{4.2 Activity and Resolution State}

Activity and resolution are separate properties. The Live, Latent, and Fresh classification describes whether a node participates in the current compilation and whether analysis has begun. It needs a resolution state to describe partial work accurately:

{\def\LTcaptype{none} 
\begin{longtable}[]{@{}
  >{\raggedright\arraybackslash}p{(\linewidth - 4\tabcolsep) * \real{0.3333}}
  >{\raggedright\arraybackslash}p{(\linewidth - 4\tabcolsep) * \real{0.3333}}
  >{\raggedright\arraybackslash}p{(\linewidth - 4\tabcolsep) * \real{0.3333}}@{}}
\toprule\noalign{}
\begin{minipage}[b]{\linewidth}\raggedright
State
\end{minipage} & \begin{minipage}[b]{\linewidth}\raggedright
Resolution
\end{minipage} & \begin{minipage}[b]{\linewidth}\raggedright
Participation and display
\end{minipage} \\
\midrule\noalign{}
\endhead
\bottomrule\noalign{}
\endlastfoot
Live, resolved & Required facts current & Eligible for the relevant compilation boundary \\
Live, pending & Some obligations unresolved & Active analysis with pending status \\
Latent, current & Retained facts still justified & Inactive, available for inspection and reuse \\
Latent, stale & One or more premises changed & Inactive, requires dependent re-analysis \\
Fresh & Parsed, not elaborated & Syntax and source diagnostics \\
\end{longtable}
}

Retaining a current subgraph can avoid repeated work. The amount saved depends on dependency structure and the validity of its boundary facts. A general \(O(\mathrm{boundary})\) reactivation bound requires an incremental algorithm and its own analysis. The editor must distinguish proved current facts from historical results whose premises need rechecking.

\subsection*{4.3 Soft Delete and Reachability}\label{soft-delete-and-reachability}
\addcontentsline{toc}{subsection}{4.3 Soft Delete and Reachability}

Our latent-preservation model retains unreachable graph structure for later inspection. Reachability can be represented per configured target, so a function may participate in the CPU build while remaining inactive for an accelerator configuration. The backend consumes the active subgraph. The language service can inspect both active and retained structures, with their provenance and freshness status.

Target unavailability must be distinguished from unreachability. An active operation requiring an unsupported capability produces a diagnostic. It becomes inactive only when the program\textquotesingle s reachability and target conditions justify excluding it.

\subsection*{4.4 Design-Time Feedback as Compilation Byproduct}\label{design-time-feedback-as-compilation-byproduct}
\addcontentsline{toc}{subsection}{4.4 Design-Time Feedback as Compilation Byproduct}

A language service could present the compiler\textquotesingle s dimensional and memory analysis through the shared graph. Hover information could show the resolved measure for a numeric expression. A representation view could show the input enclosure, its provenance, and the covering formats offered by the target. Pending obligations would remain visible before commitment.

The analyzer could also suggest a reusable lemma and dispatch its premises when the engineer accepts it. The proof application would remain attached to its source region and PSG obligations, with its current discharge status visible. An edit that invalidates a premise would trigger re-analysis. Decidable by Construction \hyperref[ref-haynes-dbc-2026]{{[}5{]}} develops this proposed authoring workflow.

Memory feedback can connect a selected placement to the escape path that required it. For a captured mutable binding, the explanation includes the shared cell and the closures that retain it. This gives the engineer a specific source location and lifetime relationship to inspect.

Layout facts also support footprint estimates. With a declared 64-byte cache line, an aligned 800-byte contiguous buffer touches \(\lceil800/64\rceil=13\) lines. An unaligned start can increase that count. Stack and arena storage can both be contiguous, and either can be evicted from cache. Loop structure alone supplies neither observed hotness nor a cache-residency guarantee.

For a cross-target transfer, BAREWire\textquotesingle s schema and layout contracts supply the field structure to be checked at both endpoints. Numeric conversion has separate range and precision obligations. IPC and network transport add their own access and communication contracts. Latency and bandwidth estimates require target declarations or measurements, with their provenance displayed alongside the estimate.

Language services still need an editing interface and incremental scheduling, but the semantic account can come from the same graph used for compilation. The engineer\textquotesingle s view should therefore show Clef dimensions and proof obligations directly, including unresolved status, rather than substituting the conventions of its current hosting language.

\section*{5. Related Work}\label{related-work}
\addcontentsline{toc}{section}{5. Related Work}

\subsection*{5.1 Units of Measure}\label{units-of-measure}
\addcontentsline{toc}{subsection}{5.1 Units of Measure}

Kennedy\textquotesingle s units-of-measure work \hyperref[ref-kennedy-2009]{{[}1{]}} established dimensional inference through integration with Hindley-Milner unification. Its implementation in F\# demonstrates the discipline in a practical language. Our DTS work retains dimensional evidence into the joint representation and placement analysis, and relates that analysis to the subsequent lowering obligations.

\subsection*{5.2 Dependent Types in F*, Idris, and Agda}\label{dependent-types-in-f-idris-and-agda}
\addcontentsline{toc}{subsection}{5.2 Dependent Types in F*, Idris, and Agda}

F* \hyperref[ref-swamy-2016]{{[}13{]}} combines dependent types and effects with solver-backed verification. Its refinements can constrain values while preserving their runtime representation. That separation influenced our treatment of \lstinline!float<newtons>! as a semantic numeric type whose concrete format is resolved independently. F* also provides a reference for incorporating automated proof discharge into an ML-family workflow.

Idris \hyperref[ref-brady-2013]{{[}14{]}} provides dependent types with a focus on practical programming. Agda \hyperref[ref-norell-2007]{{[}15{]}} is a proof assistant that also serves as a programming language. These systems and F* can encode dimensional constraints. The checking and automation guarantees depend on the selected language and fragment. F* deliberately admits solver-backed validity obligations beyond decidable fragments \hyperref[ref-fstar-smt-tutorial]{{[}16{]}}. DTS uses a principal solver for measure equations and combines it with separately identified range and proof obligations.

\subsection*{5.3 Rust Ownership and Borrow Checking}\label{rust-ownership-and-borrow-checking}
\addcontentsline{toc}{subsection}{5.3 Rust Ownership and Borrow Checking}

Rust\textquotesingle s ownership and borrowing discipline \hyperref[ref-jung-2018]{{[}17{]}} statically checks whether references remain valid. Lifetime elision covers common function signatures, with explicit lifetime relationships required where those rules leave ambiguity. The borrow checker operates on MIR with substantial type information available, and its diagnostics can explain the uses responsible for a conflict.

Our Clef design combines lifetime constraints with dimensional and platform facts in the PSG. An escaping value requires a covering lifetime, and a platform declaration determines which storage can satisfy it. A dimensionally identical computation can therefore have different concrete placements on different targets, subject to the same source contract.

\subsection*{5.4 Koka Effects and Coeffects}\label{koka-effects-and-coeffects}
\addcontentsline{toc}{subsection}{5.4 Koka Effects and Coeffects}

Koka \hyperref[ref-leijen-2014]{{[}18{]}} tracks effects in the type system, allowing implementations to specialize their handling. Our coeffect design records the contextual resource requirements that accompany an operation. Combining these with dimensional and range facts in the PSG makes it possible to relate a numeric operation\textquotesingle s meaning to its permitted placement.

\subsection*{5.5 Parametricity and Free Theorems}\label{parametricity-and-free-theorems}
\addcontentsline{toc}{subsection}{5.5 Parametricity and Free Theorems}

Reynolds \hyperref[ref-reynolds-1983]{{[}19{]}} related polymorphic types to families of related implementations. Wadler \hyperref[ref-wadler-1989]{{[}2{]}} developed the resulting free theorems. For \(g : \forall a.\; [a] \to [a]\) in the total parametric setting, the type entails \(\mathit{map}\; f \circ g = g \circ \mathit{map}\; f\) for every total function \(f\). Applying such a theorem to an effectful or partial language requires the corresponding semantic qualifications.

Kennedy\textquotesingle s dimensional invariance \hyperref[ref-kennedy-2009]{{[}1{]}} gives the specialized argument for our dimensional fragment. A function of type \lstinline!float<'d> -> float<'d> -> float<'d * 'd>! uses its abstract dimension uniformly. Source programs have no operation for case analysis on \lstinline!'d!. Dimensional relationships are checked during inference and supply the first tier of the verification architecture. Further range or lifetime properties require their own premises.

The parity extension discussed in Section 2.1 suggests a related argument. Grade involution negates odd-grade components and fixes even ones, giving a two-element group action. Establishing the corresponding invariance theorem for the combined type system requires checking the typing rules and their interaction with numeric representation. Companion work explores that extension. The physical-measure result alone does not discharge those additional obligations.

Dimensional invariance is a property of source programs. Preserving it through compiler transformations requires the separate correspondence between source and lowered operations described in Section 2.3.

\subsection*{5.6 Posit Arithmetic and Domain-Aware Representation}\label{posit-arithmetic-and-domain-aware-representation}
\addcontentsline{toc}{subsection}{5.6 Posit Arithmetic and Domain-Aware Representation}

Gustafson\textquotesingle s posit arithmetic \hyperref[ref-gustafson-yonemoto-2017]{{[}20{]}} \hyperref[ref-gustafson-2024]{{[}21{]}} provides a precision profile that varies with magnitude. The Posit Standard \hyperref[ref-posit-standard-2022]{{[}12{]}} fixes \(\mathit{es}=2\) across standard widths, while bounded-posit work investigates further constraints on the regime field. Jonnalagadda et al. \hyperref[ref-jonnalagadda-2026]{{[}22{]}} examine the hardware efficiency of those bounded configurations. Their results concern the evaluated hardware and formats, rather than a universal precision or cost advantage over IEEE arithmetic.

Posit arithmetic can be synthesized in FPGA fabric or use a declared processor instruction or accelerator. Our selection model requires the representation to cover the computation\textquotesingle s range and its realization to fit the target\textquotesingle s resources. Comparing precision then requires the actual candidate parameters and the justified enclosure. A known distribution could support a further optimization policy, but the worst-case objective in Section 2.6 has a different input: a sound bound on admitted values.

The quire also supplies a useful memory contract. Its declared width determines storage extent, and its numeric capacity constrains the permitted accumulation. DMM relates those facts to the lifetime of the operation. A smaller bounded accumulator requires evidence adequate for that configuration instead of an assumption that every quire has the standard posit32 extent.

\subsection*{5.7 MLIR and Multi-Level Compilation}\label{mlir-and-multi-level-compilation}
\addcontentsline{toc}{subsection}{5.7 MLIR and Multi-Level Compilation}

MLIR \hyperref[ref-lattner-2021]{{[}23{]}} provides extensible dialects and progressive lowering. It supplies the infrastructure for realizing the representation and placement decisions established in our PSG, subject to the preservation obligations in Section 2.3.

\subsection*{5.8 Rank Polymorphism and Shape-Indexed Types}\label{rank-polymorphism-and-shape-indexed-types}
\addcontentsline{toc}{subsection}{5.8 Rank Polymorphism and Shape-Indexed Types}

Slepak, Shivers, and Manolios develop Remora \hyperref[ref-slepak-2014]{{[}24{]}}, a rank-polymorphic array language whose type system tracks array shape as a sequence of natural-number indices. The system uses restricted dependent types to verify that rank-polymorphic lifting produces shape-consistent results, with decidable type checking and a proof of type soundness. Slepak et al. formalize rank-polymorphic type inference as constraint satisfaction over string equations \hyperref[ref-slepak-2018]{{[}25{]}}. DTS inference operates over integer linear constraints in abelian groups (Section 2.2).

Remora\textquotesingle s shape indices are sequences of natural numbers, with concatenation in a free monoid. Our physical dimensions are exponent vectors in \(\mathbb{Z}^k\). The two structures describe different properties of an array computation: how elements are arranged and what quantities those elements represent. Dynamic shapes introduce further index and existential constraints, while physical-measure equalities retain their abelian-group structure.

\section*{6. Future Work}\label{future-work}
\addcontentsline{toc}{section}{6. Future Work}

\subsection*{6.1 Formal Integration Proof}\label{formal-integration-proof}
\addcontentsline{toc}{subsection}{6.1 Formal Integration Proof}

Kennedy establishes the basis for physical-measure inference. Our remaining formal work includes the composition with memory identities, lifetime ordering, and representation constraints. A proof for the combined mechanism must specify how information passes between these domains and when their obligations are discharged. Decidability of individual components is a premise for that analysis, rather than a proof of every combined query\textquotesingle s complexity.

\subsection*{6.2 Unified Shape and Quantity Indices}\label{unified-shape-and-quantity-indices}
\addcontentsline{toc}{subsection}{6.2 Unified Shape and Quantity Indices}

A matrix of forces has both a shape, such as \(3\times4\), and an element dimension, newtons. A unified representation could retain these as independent indices and check the appropriate relationship at each operation. Shape concatenation and physical multiplication have different algebraic laws. Their combined inference requires specifying the operations and interactions before principality or complexity can be established.

\subsection*{6.3 Quantified Design-Time Feedback}\label{quantified-design-time-feedback}
\addcontentsline{toc}{subsection}{6.3 Quantified Design-Time Feedback}

The footprint calculations in Section 4.4 provide layout facts. Turning them into performance estimates requires hardware models and workload evidence. Cache hierarchy models and measured transfer costs could support quantified guidance, with the assumptions and confidence of each estimate available to the engineer.

\subsection*{6.4 Incremental Adoption Through Porting}\label{incremental-adoption-through-porting}
\addcontentsline{toc}{subsection}{6.4 Incremental Adoption Through Porting}

Porting an existing codebase would provide a practical test of the refinement workflow. Source types, library contracts, and resource conventions supply initial premises, with physical meanings introduced where the original language left them implicit. A translation can begin with partially resolved Clef code, then use diagnostics to identify which boundaries need further context.

The editor should preserve useful work while those facts are established. A sound intermediate enclosure may narrow as a guard or checked library law becomes available. The resulting representation and placement choices must be validated before the relevant code is realized. This would give the engineer a gradual path toward a complete program, with consistent partial states available for inspection.

\subsection*{6.5 Posit Hardware Co-Design and Dimensional Range Analysis}\label{posit-hardware-co-design-and-dimensional-range-analysis}
\addcontentsline{toc}{subsection}{6.5 Posit Hardware Co-Design and Dimensional Range Analysis}

Representation selection in Section 2.6 is a compile-time design. Reconfigurable targets would also permit a search among non-standard bounded-posit configurations, provided the platform describes the candidate formats and hardware resources. A finite set of exponent and regime parameters can be enumerated. Each candidate would still need coverage and precision analysis, followed by synthesis and measurement before a hardware advantage could be established.

\subsection*{6.6 Dataflow Architectures and Control-Flow/Data-Flow Partitioning}\label{dataflow-architectures-and-control-flowdata-flow-partitioning}
\addcontentsline{toc}{subsection}{6.6 Dataflow Architectures and Control-Flow/Data-Flow Partitioning}

Spatial architectures execute operations across processing elements with explicit data movement. Our PSG already represents dependencies and resource requirements that could inform such a mapping. Applying them to a spatial target requires its placement and transfer contracts to be declared and checked.

A partitioning heuristic could use dependency structure and declared resource costs to compare control-flow and spatial realizations. Regular compute regions may suit a spatial target, while other regions remain on a host processor. The quality of this choice depends on the target model and workload. A legal placement and an effective placement require different evidence.

Spatial placement also involves relations among sets of operations. AMD\textquotesingle s XDNA architecture, for example, has tiles and explicit data movement whose configuration depends on an operation group\textquotesingle s resource requirements \hyperref[ref-rico-2024]{{[}26{]}}. The companion PHG paper \hyperref[ref-haynes-phg-2026]{{[}4{]}} represents such constraints as hyperedges, extending the dimensional and coeffect relationships discussed here. Hypergraph partitioning has an established role in VLSI placement \hyperref[ref-karypis-kumar-2000]{{[}27{]}}, and MLIR-AIE \hyperref[ref-amd-xilinx-2024]{{[}28{]}} supplies relevant target infrastructure. Integrating that infrastructure with the joint proof mechanism remains a separate implementation and validation task.

\subsection*{6.7 Extending Continuation and Reduction Coverage}\label{extending-continuation-and-reduction-coverage}
\addcontentsline{toc}{subsection}{6.7 Extending Continuation and Reduction Coverage}

The graph construction described in Section 4.1 supplies the current architectural basis for continuations and joint reduction. The remaining task is to extend and validate the admitted source forms. A continuation contract must identify which computation is resumed, which environment it retains, and whether its uses preserve sharing and storage lifetime. A reduction rule must state its participants, retained resources and permitted duplication, with a preservation argument for the resulting graph.

Computation expressions provide structured entry points for investigating these contracts. Broader scheduling and parallel execution require evidence about effects, sharing, resource availability and progress. The existence of frame and dispatch construction does not supply a general evaluator or a confluence proof for arbitrary interaction rules. The companion PHG \hyperref[ref-haynes-phg-2026]{{[}4{]}} and FPS \hyperref[ref-haynes-fps-2026]{{[}6{]}} papers develop the joint-relation and preservation questions.

\subsection*{6.8 Formal Verification Integration}\label{formal-verification-integration}
\addcontentsline{toc}{subsection}{6.8 Formal Verification Integration}

A suitable first integration case is a buffer access: the proof must connect the PSG\textquotesingle s index and BAREWire layout facts to the access that will execute. The canonical mechanism retains that obligation beside its computation. During development, the external ledger described in Section 2.3 compares the PSG and lowered obligations. Its removal requires evidence that the graph preserves their identities, premises and results across the boundary.

Measure equalities use the dimensional group solver. Range and layout obligations may use decidable theories such as QF\_LIA or QF\_BV \hyperref[ref-barrett-2010]{{[}29{]}} and be dispatched to cvc5. A timeout or unsupported obligation remains unresolved and blocks a dependent commitment. The Decidable by Construction \hyperref[ref-haynes-dbc-2026]{{[}5{]}} and Fixed-Point Scaffolding \hyperref[ref-haynes-fps-2026]{{[}6{]}} papers develop the higher-tier lemma and relational mechanisms.

A certified lowering pass carries a checked proof transformation. Other passes require a re-dispatch sufficient to establish the lower operation\textquotesingle s precondition from retained premises. A proof example should expose those premises, identify the source and target operations, and state which checker accepts the evidence. This would let an engineer read the guarantee before inspecting its machine-generated derivation.

\subsection*{6.9 Information Accrual and Deferred Optimization}\label{information-accrual-and-deferred-optimization}
\addcontentsline{toc}{subsection}{6.9 Information Accrual and Deferred Optimization}

A generic function may be dimensionally typed before a caller supplies a target. Platform quotations can later settle its available formats and memory layout. A decision becomes justified when its premises are established in the current context:

\[
\operatorname{Premises}(D) \subseteq \operatorname{EstablishedFacts}(\text{current context}).
\]

It must settle before a consumer requires it. Source dimensions, representation and placement therefore resolve in the PSG before witnessing. Later stages may release earlier type structure once its consumers are satisfied. Actual call frequencies still require profiling or a justified workload model, and source edits can invalidate earlier conclusions as described in Section 4.1.

Deferred inference concerns the compiler\textquotesingle s knowledge; deferred evaluation concerns when the program demands a computation. The negative and fractional type proposal \hyperref[ref-haynes-nft-2026]{{[}30{]}} adds pairing and cancellation obligations to investigate within this architecture. The ADM proposal \hyperref[ref-haynes-adm-2026]{{[}10{]}} uses changing probabilistic evidence to select candidates. Both retain the requirement that a candidate satisfy the source and target contracts before it is committed.

\subsection*{6.10 Implications for Numerically Disciplined Machine Learning}\label{implications-for-numerically-disciplined-machine-learning}
\addcontentsline{toc}{subsection}{6.10 Implications for Numerically Disciplined Machine Learning}

\textbf{Dimensional algebra under differentiation.} The dimensional algebra is closed under differentiation. If \(f : \mathbb{R}^{\langle d_1 \rangle} \to \mathbb{R}^{\langle d_2 \rangle}\), where \(\langle d \rangle\) denotes the dimensional annotation, then:

\[
\frac{\partial f}{\partial x} : \mathbb{R}^{\langle d_2 \cdot d_1^{-1} \rangle}
\]

The gradient of a loss function with dimension \(\langle \text{loss} \rangle\) with respect to a parameter with dimension \(\langle d \rangle\) carries dimension \(\langle \text{loss} \cdot d^{-1} \rangle\). This property follows from the abelian group structure: differentiation is division in the dimensional algebra at the level of each partial derivative, and division is closed in \(\mathbb{Z}^n\). An automatic-differentiation transformation would need to generate the appropriate dimensional constraints for its derivative rules. Once generated, those equalities use the same abelian-group unifier as the forward computation.

For a vector-valued function, Jacobian entries may have different physical dimensions. If \(f : \mathbb{R}^{\langle d^{\mathrm{in}}_1\rangle}\times\cdots\times\mathbb{R}^{\langle d^{\mathrm{in}}_n\rangle}\to\mathbb{R}^{\langle d^{\mathrm{out}}_1\rangle}\times\cdots\times\mathbb{R}^{\langle d^{\mathrm{out}}_m\rangle}\), entry \((i,j)\) has dimension \(\langle d^{\mathrm{out}}_i\cdot(d^{\mathrm{in}}_j)^{-1}\rangle\). The representation must retain those distinctions when the entries are combined. For a uniformly dimensionless model, every entry has the same dimensional identity.

For a physics-informed model, this would check the dimensions of each loss derivative. Components measured in newtons per meter and joules per second require an explicit dimensionally valid relationship before they can be combined. Adding force and energy residuals to a single scalar loss likewise requires appropriate normalization or weighting. These are structural checks on the differentiated computation, independent of the model\textquotesingle s statistical accuracy.

\textbf{Dimensions of a probabilistic metric.} Our ADM proposal \hyperref[ref-haynes-adm-2026]{{[}10{]}} develops a restricted natural-gradient construction using the score-velocity geometry of Karwowski, Nielsen, and Staton \hyperref[ref-karwowski-geometry-2026]{{[}31{]}}. Its dimensional admission belongs to the same derivative rules. Let parameter \(\theta_i\) have dimension \(U_i\), output coordinate \(y_a\) have dimension \(W_a\), score component \(S_i\) be the parameter derivative of the log density under a fixed reference measure, and \(V_{ai}=\partial y_a/\partial\theta_i\). For an output metric \(h\) and a nonnegative weight \(\kappa\), the required component dimensions are

\[
[S_i]=U_i^{-1},\qquad [V_{ai}]=W_aU_i^{-1},\qquad
[\kappa h_{ab}]=W_a^{-1}W_b^{-1}.
\]

Here brackets denote physical dimension. Both the score product \(S_iS_j\) and each weighted velocity contribution \(\kappa h_{ab}V_{ai}V_{bj}\) then have dimension \(U_i^{-1}U_j^{-1}\). One admissible choice uses fixed output reference scales \(r_a\) and \(h=\sum_a(dy_a/r_a)^2\), with dimensionless \(\kappa\). If \(h\) instead measures squared length, \(\kappa\) must carry inverse squared length. The scales and relative weight specify the optimizer\textquotesingle s geometry; dimensional compatibility alone does not denote values. Heterogeneous outputs require the corresponding coordinate-specific factors before their contributions can be added.

Fixed parameter scales \(s_i\) similarly give dimensionless coordinates \(z_i=\theta_i/s_i\). The metric and loss differential must both be transported to those coordinates. The resulting natural-gradient direction has the loss\textquotesingle s units, so an update in \(z\) requires a step coefficient with their inverse; dimensionless loss admits a dimensionless coefficient. These checks concern the declared metric and coordinate interpretation. Existence of the metric, its kernel, and the statistical accuracy of its estimated entries retain the separate obligations developed in ADM and DBC \hyperref[ref-haynes-dbc-2026]{{[}5{]}}.

\textbf{Forward-mode differentiation as a coeffect property.} Baydin et al. \hyperref[ref-baydin-2022]{{[}32{]}} construct an unbiased gradient estimator using a forward directional derivative and a sampled tangent. The estimator avoids the conventional backward activation tape. Its sampling variance and computational tradeoffs must be evaluated for the workload. Reverse-mode provides the full gradient through its derivative rules, subject to the numerical behavior of their implementation.

Flügel et al. \hyperref[ref-flugel-2026]{{[}33{]}} develop a multi-tangent extension using orthogonal projection. Its additional state and computation requirements are discussed below.

Within DMM, reverse-mode\textquotesingle s retained activations are contextual resources whose lifetimes span the relevant backward uses. Checkpointing and recomputation can reduce that storage at a computational cost. Forward-mode carries primal and tangent state through the computation and avoids the conventional backward activation tape, but the size of that live state depends on the model and tangent strategy.

{\def\LTcaptype{none} 
\begin{longtable}[]{@{}
  >{\raggedright\arraybackslash}p{(\linewidth - 4\tabcolsep) * \real{0.3333}}
  >{\raggedright\arraybackslash}p{(\linewidth - 4\tabcolsep) * \real{0.3333}}
  >{\raggedright\arraybackslash}p{(\linewidth - 4\tabcolsep) * \real{0.3333}}@{}}
\toprule\noalign{}
\begin{minipage}[b]{\linewidth}\raggedright
AD mode
\end{minipage} & \begin{minipage}[b]{\linewidth}\raggedright
Retained state
\end{minipage} & \begin{minipage}[b]{\linewidth}\raggedright
Lifetime question
\end{minipage} \\
\midrule\noalign{}
\endhead
\bottomrule\noalign{}
\endlastfoot
Reverse-mode & Activations needed by backward uses, subject to recomputation & Which forward values must survive to those uses? \\
Forward-mode estimator & Live primal and tangent state & Which values are retained by later layers, outputs, or callbacks? \\
\end{longtable}
}

Eliminating the tape removes a substantial source of retention. The remaining state still requires escape analysis. Layer outputs and skip connections can outlive a local scope, and a permitted stack allocation must also fit the target\textquotesingle s capacity. DMM must therefore check the live state and its uses for the selected evaluation strategy.

The proposed probabilistic geometry adds consumers of sampled choices, density contributions and directional state. DMM would account for their actual evaluation strategy: streaming accumulation, sufficient retained statistics, checkpointed recomputation or an explicit trace record. A mathematical trace space does not itself require a complete runtime log. Replacing retained state by a smaller representation needs a contract covering every later score, metric or recovery use. Replay additionally requires the relevant random-generator state, environments and effects to determine the claimed result. A seed alone supplies no general recovery guarantee, and geometry sufficient for an update need not retain enough information to reconstruct an execution.

Quire-backed product accumulation can reduce rounding within the derivative calculation when the target supports it and the accumulation bound is established. Other derivative operations and the final readout retain their own rounding behavior. The resulting gradient therefore has separate dimensional, memory, statistical, and numerical obligations.

\textbf{Multi-tangent extension and the \(k/n\) ratio as a design-time property.} Flügel et al. \hyperref[ref-flugel-2026]{{[}33{]}} investigate forward-gradient estimators using multiple tangents. In Euclidean coordinates with a declared reference scaling for physically heterogeneous parameters, and with linearly independent columns in \(\mathbf{V}\), the orthogonal projection onto their span is:

\[
P_U(\nabla f)=\mathbf{V}(\mathbf{V}^{\top}\mathbf{V})^{-1}\mathbf{V}^{\top}\nabla f.
\]

It equals the gradient when that gradient lies in \(U\). The directional observations \(\mathbf{V}^{\top}\nabla f\) leave any component orthogonal to \(U\) undetermined, so they alone cannot certify the omitted residual. Statistical quality depends on how the tangents are chosen and on the estimator\textquotesingle s assumptions. The parameter-space size \(n\) comes from tensor shape and model structure, rather than physical dimensions such as meters. If \(n\) and the tangent count \(k\) are known, the compiler could display their ratio and the associated resource requirements. An expected quality estimate needs its probabilistic premises and a distinction from an individual sampled result.

Multiple tangents increase live state or repeated work according to the evaluation strategy. The Gram matrix adds storage, and solving its system has conditioning and rounding requirements. Exact accumulation of some matrix products alone does not make matrix inversion exact.

\textbf{Representation selection for neural network values.} Training statistics can suggest useful candidate formats or rescalings. A sound enclosure supplies the coverage requirement, while a distribution describes how often values occur within its support. These are different inputs. Our worst-case objective in Section 2.6 scores covering candidates over the enclosure. A future distribution-sensitive policy would require an explicit objective and provenance for its evidence.

Bounded-posit configurations offer parameters for exploring this tradeoff, including exponent and regime choices discussed by Gustafson \hyperref[ref-gustafson-2024]{{[}21{]}}. A bias shift or rescaling can change where precision is concentrated, with conversions and recomputed range obligations at the boundaries. The usefulness of a configuration for a particular model requires evaluation. Its legality still depends on the numeric contract.

\textbf{Physics-informed loss term verification.} Physics-informed neural networks \hyperref[ref-raissi-2019]{{[}34{]}} encode physical laws as differentiable loss terms. For a residual \(F-ma\), dimensional inference verifies that force, mass, and acceleration compose to a common unit before subtraction. Squaring the residual gives a squared-force dimension, which must be reconciled with any other terms in the loss. This establishes a property of the loss function\textquotesingle s structure. Accuracy of the trained model remains a separate empirical and mathematical question.

\section*{7. Semantic Continuity}\label{semantic-continuity}
\addcontentsline{toc}{section}{7. Semantic Continuity}

An engineer should be able to begin with a numeric relationship and add the context needed to realize it on a target. The dimensional algebra gives that relationship a stable type identity. Range and lifetime constraints can then refine the permissible representations and placements, with a specific obligation at each point where a decision becomes necessary.

Our immediate implementation work tests that continuity at primitive operations: measures must survive elaboration, declared memory facts must reach the PSG, and the corresponding lowered operations must be checked against the same premises. The temporary verification ledger provides an independent comparison while the proof-carrying mechanism is established. The eventual editing experience depends on that account being accurate enough to explain both a completed decision and a fact still awaiting context.

\section*{Acknowledgments}\label{acknowledgments}
\addcontentsline{toc}{section}{Acknowledgments}

This paper owes a particular debt to John L. Gustafson, whose detailed correspondence on posit arithmetic, bounded posit parameterization, and domain-specific precision tuning shaped how the author thinks about representation selection. The treatment of asymmetric b-posit configurations and hardware reuse reflects his influence directly.

Don Syme\textquotesingle s language-design work has informed Clef\textquotesingle s development. His feedback on this manuscript sharpened the framing of dimensional persistence and the relationship between annotation provenance and compilation-stage decisions.

Paul Snively provided early guidance on verification reference materials that opened a line of investigation the author would not have pursued otherwise. The formal verification aspects of the Fidelity framework research bear his mark. Martin Coll\textquotesingle s interaction-net work and his ongoing engagement with the Fidelity project have been a consistent source of both technical insight and encouragement.

\section*{Software Availability}\label{software-availability}
\addcontentsline{toc}{section}{Software Availability}

The Clef language, Composer compiler, and supporting libraries described in this paper are developed under the Fidelity Framework project. Source repositories are available at \url{https://github.com/FidelityFramework}. The language specification, compiler documentation, and ongoing project research are available at \url{https://clef-lang.com}. Central components of the framework are dual-licensed. Terms are detailed in each repository. All components referenced in this paper, including the DTS inference engine, escape analysis pipeline, and BAREWire interchange protocol, are under active development.

The implementation observations in Sections 2.2 and 4.1 refer to Clef revision \lstinline!7ad1daf23364!: \lstinline!MemberInferenceCases.fs! records the ordinary and inline member cases, and \nolinkurl{Baker/Ingredients/Continuations.fs} supplies the frame and dispatch construction. The demand and capture account follows \lstinline!expressions.md! and \lstinline!closure-representation.md! at specification revision \lstinline!4d5c445331b3!. These sources establish the cited contracts and mechanisms; this paper reports no new execution of the compiler\textquotesingle s complete acceptance suite.

\section*{References}\label{references}
\addcontentsline{toc}{section}{References}

\leavevmode\protect\phantomsection\label{ref-kennedy-2009}{}{[}1{]} Kennedy, A. (2009). Types for units-of-measure: Theory and practice. In \emph{Central European Functional Programming School}, LNCS 6299. Springer.

\leavevmode\protect\phantomsection\label{ref-wadler-1989}{}{[}2{]} Wadler, P. (1989). Theorems for free! In \emph{Proceedings of FPCA}, pages 347-359. ACM.

\leavevmode\protect\phantomsection\label{ref-sarkar-2004}{}{[}3{]} Sarkar, D., Waddell, O., and Dybvig, R. K. (2004). A nanopass infrastructure for compiler education. In \emph{Proceedings of the Ninth ACM SIGPLAN International Conference on Functional Programming} (ICFP \textquotesingle04), pages 201-212. ACM.

\leavevmode\protect\phantomsection\label{ref-haynes-phg-2026}{}{[}4{]} Haynes, H. (2026). The program hypergraph: Multi-way relational structure for geometric algebra, spatial compute, and physics-aware compilation. \emph{arXiv preprint arXiv:2603.17627}.

\leavevmode\protect\phantomsection\label{ref-haynes-dbc-2026}{}{[}5{]} Haynes, H. (2026). Decidable by construction: Design-time verification for trustworthy AI. \href{https://arxiv.org/abs/2603.25414}{arXiv:2603.25414}.

\leavevmode\protect\phantomsection\label{ref-haynes-fps-2026}{}{[}6{]} Haynes, H. (2026). Fixed-point scaffolding in the Clef programming language. \href{https://arxiv.org/abs/2606.02854}{arXiv:2606.02854}.

\leavevmode\protect\phantomsection\label{ref-harper-1990}{}{[}7{]} Harper, R., Mitchell, J. C., and Moggi, E. (1990). Higher-order modules and the phase distinction. In \emph{Proceedings of the 17th ACM SIGPLAN-SIGACT Symposium on Principles of Programming Languages} (POPL), pages 341-354. ACM.

\leavevmode\protect\phantomsection\label{ref-nvidia-floating-point}{}{[}8{]} NVIDIA. \emph{Floating Point and IEEE 754 Compliance for NVIDIA GPUs}. CUDA documentation. \href{https://docs.nvidia.com/cuda/floating-point/index.html}{Guide}.

\leavevmode\protect\phantomsection\label{ref-arm-compiler-guide}{}{[}9{]} Arm. \emph{Arm Compiler for Embedded User Guide}. \href{https://documentation-service.arm.com/static/641d6ff4ac798355e7426c4c\#page=66}{Guide, p. 66}.

\leavevmode\protect\phantomsection\label{ref-haynes-adm-2026}{}{[}10{]} Haynes, H. (2026). Adaptive domain models: Bayesian evolution, warm rotation, and principled training for geometric and neuromorphic AI. \href{https://arxiv.org/abs/2603.18104}{arXiv:2603.18104}.

\leavevmode\protect\phantomsection\label{ref-petricek-2014}{}{[}11{]} Petricek, T., Orchard, D., and Mycroft, A. (2014). Coeffects: A calculus of context-dependent computation. In \emph{Proceedings of ICFP}.

\leavevmode\protect\phantomsection\label{ref-posit-standard-2022}{}{[}12{]} Posit Working Group. (2022). Standard for posit arithmetic (2022). \url{https://posithub.org/docs/posit_standard-2.pdf}

\leavevmode\protect\phantomsection\label{ref-swamy-2016}{}{[}13{]} Swamy, N., Hriţcu, C., Keller, C., Rastogi, A., Delignat-Lavaud, A., Forest, S., Bhargavan, K., Fournet, C., Strub, P.-Y., Kohlweiss, M., Zinzindohoué, J.-K., and Zanella-Béguelin, S. (2016). Dependent types and multi-monadic effects in F\emph{. In }Proceedings of POPL*.

\leavevmode\protect\phantomsection\label{ref-brady-2013}{}{[}14{]} Brady, E. (2013). Idris, a general-purpose dependently typed programming language: Design and implementation. \emph{Journal of Functional Programming}, 23(5).

\leavevmode\protect\phantomsection\label{ref-norell-2007}{}{[}15{]} Norell, U. (2007). \emph{Towards a practical programming language based on dependent type theory}. PhD thesis, Chalmers University of Technology.

\leavevmode\protect\phantomsection\label{ref-fstar-smt-tutorial}{}{[}16{]} F* contributors. Interfacing with an SMT solver. \emph{Proof-Oriented Programming in F*}. \href{https://fstar-lang.org/tutorial/book/part1/part1_prop_assertions.html}{Tutorial}.

\leavevmode\protect\phantomsection\label{ref-jung-2018}{}{[}17{]} Jung, R., Jourdan, J.-H., Krebbers, R., and Dreyer, D. (2018). RustBelt: Securing the foundations of the Rust programming language. In \emph{Proceedings of POPL}.

\leavevmode\protect\phantomsection\label{ref-leijen-2014}{}{[}18{]} Leijen, D. (2014). Koka: Programming with row polymorphic effect types. In \emph{Proceedings of the 5th Workshop on Mathematically Structured Functional Programming} (MSFP).

\leavevmode\protect\phantomsection\label{ref-reynolds-1983}{}{[}19{]} Reynolds, J. C. (1983). Types, abstraction and parametric polymorphism. In \emph{Information Processing 83}, pages 513-523. Elsevier.

\leavevmode\protect\phantomsection\label{ref-gustafson-yonemoto-2017}{}{[}20{]} Gustafson, J. L., and Yonemoto, I. T. (2017). Beating floating point at its own game: Posit arithmetic. \emph{Supercomputing Frontiers and Innovations}, 4(2).

\leavevmode\protect\phantomsection\label{ref-gustafson-2024}{}{[}21{]} Gustafson, J. L. (2024). \emph{Every Bit Counts: Posit Computing}. Chapman \& Hall/CRC.

\leavevmode\protect\phantomsection\label{ref-jonnalagadda-2026}{}{[}22{]} Jonnalagadda, A. A., Thotli, R., and Gustafson, J. L. (2026). Closing the gap between float and posit hardware efficiency. In \emph{Conference on Next Generation Arithmetic}. arXiv preprint arXiv:2603.01615.

\leavevmode\protect\phantomsection\label{ref-lattner-2021}{}{[}23{]} Lattner, C., Amini, M., Bondhugula, U., Cohen, A., Davis, A., Pienaar, J., Riddle, R., Shpeisman, T., Vasilache, N., and Zinenko, O. (2021). MLIR: Scaling compiler infrastructure for domain specific computation. In \emph{Proceedings of CGO}.

\leavevmode\protect\phantomsection\label{ref-slepak-2014}{}{[}24{]} Slepak, J., Shivers, O., and Manolios, P. (2014). An array-oriented language with static rank polymorphism. In \emph{Proceedings of ESOP}, LNCS 8410, pages 27-46. Springer.

\leavevmode\protect\phantomsection\label{ref-slepak-2018}{}{[}25{]} Slepak, J., Manolios, P., and Shivers, O. (2018). Rank polymorphism viewed as a constraint problem. In \emph{Proceedings of the 5th ACM SIGPLAN International Workshop on Libraries, Languages and Compilers for Array Programming} (\href{mailto:ARRAY@PLDI}{\nolinkurl{ARRAY@PLDI}}).

\leavevmode\protect\phantomsection\label{ref-rico-2024}{}{[}26{]} Rico, A., Pareek, S., Cabezas, J., Clarke, D., et al. (2024). AMD XDNA NPU in Ryzen AI processors. \emph{IEEE Micro}, 44(6), 73-83.

\leavevmode\protect\phantomsection\label{ref-karypis-kumar-2000}{}{[}27{]} Karypis, G., and Kumar, V. (2000). Multilevel \(k\)-way hypergraph partitioning. \emph{VLSI Design}, 11(3), 285-300.

\leavevmode\protect\phantomsection\label{ref-amd-xilinx-2024}{}{[}28{]} AMD/Xilinx. (2024). MLIR-AIE: An MLIR-based toolchain for AMD AI engines. \url{https://github.com/Xilinx/mlir-aie}

\leavevmode\protect\phantomsection\label{ref-barrett-2010}{}{[}29{]} Barrett, C., Stump, A., and Tinelli, C. (2010). The SMT-LIB standard: Version 2.0. In \emph{Proceedings of the 8th International Workshop on Satisfiability Modulo Theories} (SMT).

\leavevmode\protect\phantomsection\label{ref-haynes-nft-2026}{}{[}30{]} Haynes, H. (2026). Negative and fractional types in the Fidelity Framework: A proposed operational core for reversibility and resource pairing. \href{https://arxiv.org/abs/2606.04352}{arXiv:2606.04352}.

\leavevmode\protect\phantomsection\label{ref-karwowski-geometry-2026}{}{[}31{]} Karwowski, J., Nielsen, F., and Staton, S. (2026). Compositional information geometry of probabilistic programs. \href{https://arxiv.org/abs/2610.17886v1}{arXiv:2610.17886v1}.

\leavevmode\protect\phantomsection\label{ref-baydin-2022}{}{[}32{]} Baydin, A. G., Pearlmutter, B. A., Syme, D., Wood, F., and Torr, P. (2022). Gradients without backpropagation. \emph{arXiv preprint arXiv:2202.08587}.

\leavevmode\protect\phantomsection\label{ref-flugel-2026}{}{[}33{]} Flügel, K., Coquelin, D., Weiel, M., Debus, C., Streit, A., and Götz, M. (2026). Beyond backpropagation: Optimization with multi-tangent forward gradients. \emph{arXiv preprint arXiv:2410.17764}. Revised January 2026.

\leavevmode\protect\phantomsection\label{ref-raissi-2019}{}{[}34{]} Raissi, M., Perdikaris, P., and Karniadakis, G. E. (2019). Physics-informed neural networks: A deep learning framework for solving forward and inverse problems involving nonlinear partial differential equations. \emph{Journal of Computational Physics}, 378, 686-707.

\section*{Appendix A. DTS Inference Example}\label{appendix-a.-dts-inference-example}
\addcontentsline{toc}{section}{Appendix A. DTS Inference Example}

Consider an unannotated function with the gravitational constant supplied as an argument:

\begin{lstlisting}
let computeForce gravitationalConstant mass1 mass2 distance =
    gravitationalConstant * mass1 * mass2 / (distance * distance)
\end{lstlisting}

Assign dimension variables \(d_g\), \(d_{m1}\), \(d_{m2}\), and \(d_r\) to the arguments. In additive exponent-vector notation, multiplication and division give:

\[
d_{\mathrm{return}}=d_g+d_{m1}+d_{m2}-2d_r.
\]

The function can remain dimensionally polymorphic. At a floating specialization, supplying arguments with dimensions \(\mathrm{m}^3\mathrm{kg}^{-1}\mathrm{s}^{-2}\), \(\mathrm{kg}\), \(\mathrm{kg}\), and \(\mathrm{m}\) resolves the return dimension as:

\[
(3,-2,-1)+(0,0,1)+(0,0,1)-2(1,0,0)=(1,-2,1),
\]

using length, time, and mass as the coordinate order. The result is force, \(\mathrm{kg}\,\mathrm{m}\,\mathrm{s}^{-2}\). A call can obtain the constant\textquotesingle s dimension from an explicit measured literal or a checked library declaration.

A bare literal \lstinline!6.674e-11! is dimensionless. Its numerical resemblance to the gravitational constant supplies no unit. Replacing the parameter with that bare literal would give a different dimensional relationship, and a force annotation at the result would expose the mismatch. The inference derives relationships from declared and inferred premises. It does not recognize physical facts from a decimal spelling.

\section*{Appendix B. Escape Analysis and Restructuring Example}\label{appendix-b.-escape-analysis-and-restructuring-example}
\addcontentsline{toc}{section}{Appendix B. Escape Analysis and Restructuring Example}

A sensor-processing function can return a newly produced span:

\begin{lstlisting}
let processReadings (sensors: Span<float<volts>>) =
    let readings = sensors |> Span.map (fun s -> s * calibrationFactor)
    let summary = summarize readings
    (readings, summary)
\end{lstlisting}

For this example, \lstinline!calibrationFactor! is dimensionless, and the span operations have their ordinary mapping contracts. The returned view requires its backing storage to remain valid for the caller\textquotesingle s uses. If that storage was tentatively placed in the callee\textquotesingle s stack frame, the return creates a lifetime obligation. A covering caller region or another permitted storage class can satisfy it. The escape itself does not mandate an arena.

An editor could display the return path and offer the following restructurings, subject to checking the changed program.

\textbf{Alternative 1: Caller-provided buffer.}

\begin{lstlisting}
let processReadings (sensors: Span<float<volts>>)
                    (output: Span<float<volts>>) =
    sensors |> Span.mapInto output (fun s -> s * calibrationFactor)
    summarize output
\end{lstlisting}

The callee can reuse the output buffer if its capacity and access permissions are adequate. The caller owns the placement and must establish a lifetime covering all uses. Eliminating an allocation in this function does not imply that creating the caller\textquotesingle s buffer has zero cost.

\textbf{Alternative 2: Continuation style.}

\begin{lstlisting}
let processReadings (sensors: Span<float<volts>>)
                    (k: Span<float<volts>> -> Summary -> 'a) =
    let readings = sensors |> Span.map (fun s -> s * calibrationFactor)
    k readings (summarize readings)
\end{lstlisting}

The callback executes before the callee returns, but can still retain \lstinline!readings! or return a reference to it. Stack placement requires evidence that neither occurs beyond the backing storage\textquotesingle s lifetime. A checked non-retention contract or analysis of the instantiated continuation supplies that evidence.

\textbf{Alternative 3: Declared region.}

A platform quotation can supply a region whose lifetime covers the returned span. The mapping operation allocates through the permitted resource, and the compiler checks capacity, alignment, and release boundaries. An annotation documenting the intended region remains subject to those obligations when a dependency changes.

\section*{Appendix C. Representation Selection with Posit Arithmetic}\label{appendix-c.-representation-selection-with-posit-arithmetic}
\addcontentsline{toc}{section}{Appendix C. Representation Selection with Posit Arithmetic}

Consider the force computation with a measured gravitational constant:

\begin{lstlisting}
let computeForce (m1: float<kg>) (m2: float<kg>) (r: float<m>)
    : float<newtons> =
    let g = 6.674e-11<m^3 * kg^-1 * s^-2>
    g * m1 * m2 / (r * r)
\end{lstlisting}

For the mathematical expression and declared positive input intervals \(m_1,m_2\in[10^{22},10^{30}]\) kg and \(r\in[10^6,10^{11}]\) meters, interval propagation gives:

\[
F\in\left[\frac{6.674\times10^{-11}\times10^{44}}{10^{22}},
          \frac{6.674\times10^{-11}\times10^{60}}{10^{12}}\right]
 =[6.674\times10^{11},\;6.674\times10^{37}]\ \mathrm{newtons}.
\]

A declared IEEE binary64 capability covers these finite magnitudes. On an FPGA, the arithmetic circuit is synthesized for the selected representation. Standard posit32 cannot cover this result enclosure, whose upper endpoint exceeds \(2^{120}\approx1.3\times10^{36}\), so selection must consider a covering format that fits the available fabric resources. Intermediate operations also require coverage. Rounding behavior must be included in the realized operation contracts rather than inferred from this ideal-arithmetic interval alone.

An editor could show the outcome for each declared target separately:

\begin{lstlisting}
Target: CPU with declared IEEE binary64 capability
computeForce: float<kg> -> float<kg> -> float<m> -> float<newtons>
  Range: [6.674e11, 6.674e37]
  Result range: covered with the processor's accelerator resources
\end{lstlisting}

\begin{lstlisting}
Target: FPGA
computeForce: float<kg> -> float<kg> -> float<m> -> float<newtons>
  Range: [6.674e11, 6.674e37]
  posit32: insufficient range
  Required: a covering representation within the fabric's resources
\end{lstlisting}

Smaller justified input ranges may make posit32 eligible, but that conclusion requires recomputing every affected enclosure. That is one reason we are building Composer as a differential compiler. Recomputation is scoped to the affected PSG regions to keep these choices responsive as the engineer refines the program.

If a required boundary fixes posit32 where it\textquotesingle s \emph{not} eligible, that failure to cover the range is a design-time error diagnostic. The decimal spelling of the constant and the names of the parameters supply no implicit astrophysical bounds. Where a transfer between posit32 and binary64 is required, BAREWire supplies the applicable layout and communication contract.

Numeric fidelity is checked separately from BAREWire\textquotesingle s layout and communication contracts: a correct transfer does not by itself establish that conversion preserved the numeric value. Every finite standard posit32 value has at most 28 significant binary digits and an exponent within binary64\textquotesingle s range, so those finite values admit an exact conversion to binary64. Range inclusion alone would not prove exactness. The NaR case needs the declared exceptional-value policy, and conversion in the reverse direction requires its own rounding and coverage analysis.

Representation adequacy, layout compatibility, and transfer fidelity therefore remain separate obligations attached to the corresponding operations. The witness for a boundary must preserve the evidence for each obligation it consumes.
\end{document}